\documentclass[final,5p,times,twocolumn,authoryear]{elsarticle}

\usepackage{cite}
\usepackage{amsmath,amssymb,epsfig}
\usepackage[dvipsnames]{xcolor}
\usepackage{mathrsfs}
\usepackage{color}
\usepackage{aas_macros}
\usepackage{textcomp}
\usepackage{ wasysym }
\usepackage{fontawesome}
\usepackage{dirtytalk}
\usepackage{hyperref}

\newcommand{\beq}{\begin{equation}}
\newcommand{\eeq}{\end{equation}}

\usepackage{multicol}

\journal{Astronomy and Computing}

\begin{document}
\begin{frontmatter}
\title{MADLens, a python package for fast and differentiable non-Gaussian lensing simulations}

\author[a,b]{Vanessa B\"ohm \corref{cor1} }
\ead{vboehm@berkeley.edu}
\author[a]{Yu Feng,}
\author[a]{Max E. Lee,}
\author[a]{Biwei Dai}

\address[a]{Berkeley Center for Cosmological Physics, University of California, Berkeley, CA 94720, USA}
\address[b]{Lawrence Berkeley National Lab, 1 Cyclotron Road, Berkeley, CA 94720, USA}

\cortext[cor1]{Corresponding author}

\begin{abstract}
We present MADLens a python package for producing non-Gaussian lensing convergence maps at arbitrary source redshifts with unprecedented precision. MADLens is designed to achieve high accuracy while keeping computational costs as low as possible. A MADLens simulation with only $256^3$ particles produces convergence maps whose power agree with theoretical lensing power spectra up to $L{=}10000$ within the accuracy limits of HaloFit. This is made possible by a combination of a highly parallelizable particle-mesh algorithm, a sub-evolution scheme in the lensing projection, and a machine-learning inspired sharpening step. Further, MADLens is fully differentiable with respect to the initial conditions of the underlying particle-mesh simulations and a number of cosmological parameters. These properties allow MADLens to be used as a forward model in Bayesian inference algorithms that require optimization or derivative-aided sampling. Another use case for MADLens is the production of large, high resolution simulation sets as they are required for training novel deep-learning-based lensing analysis tools. We make the MADLens package publicly available under a Creative Commons License~\href{https://github.com/VMBoehm/MADLens}{\faGithub}.
\end{abstract}


\end{frontmatter}

\section{Introduction}
Measurements of the weak cosmic shear signal will be among the major experimental drivers for advancing cosmology in the next decade. Next generation surveys such as \textit{LSST}~\citep{LSST}, the \textit{Roman Space Telescope}~\citep{WFIRST} and the \textit{EUCLID} satellite~\citep{EUCLID} will provide an unprecedented amount of high resolution weak cosmic shear data, which creates a demand for novel data analysis and modeling techniques. The weak cosmic shear signal is sensitive to the evolution of matter clustering over several orders of scales, ranging from well within the linear to the highly non-linear regime. Tomographic lensing measurements are sensitive to the total matter content, $\Omega_{m0}$, the amplitude of clustering, $\sigma_8$, and the time evolution of clustering, which allows to constrain dark energy~\citep{Hu02,Huterer02,Song04} and the sum of neutrino masses. Weak cosmic shear measurements can further be used to test general relativity~\citep{Heavens07,Schmidt08,Deraps15}.

Traditional lensing analyses mostly rely on two-point statistics or related observables~\citep{Kitching11,Heymans13,Kitching14,Alsing16,Hildebrandt17,Troxel18,Taylor19, Hikage19} to extract cosmological information. However, since the lensing convergence field is inherently and significantly non-Gaussian, two-point statistics do not exploit its full information content. In fact, a long list of studies have shown that non-Gaussian summary statistics, such as higher order correlation functions~\citep{Pen03,Takada03,Jarvis04,Sembolini11,Fu14,Coulton19} or peak statistics~\citep{Jain00,Dietrich10,Maturi11,Marian12, Pires12, Petri13, Cardone13,Lin15, Liu15,LiuX15,Kacprzak16,Peel17,Liu19} can break parameter degeneracies that cumber power spectrum analyses and lead to significantly tighter constraints. While inference from these summaries offers improvements over power spectra analyses, their choice is somewhat ad hoc and the question of how to best extract cosmological information from non-Gaussian lensing scales is still an active field of research.

A number of works have recently suggested machine-learning tools for identifying informative summaries~\citep{Gupta18,Ribli19} and even successfully applied them to real data~\citep{Fluri19}. Machine learning methods require a large amount of training data. If applied correctly, these methods \textit{learn} which features in the lensing map are most informative about cosmological parameters. A recent study finds that these models are mostly sensitive to extreme values in the lensing field~\citep{Zotilla20}. This underlines the importance of training data that accurately mimics real cosmic shear data and its dependence on cosmological parameters down to very small scales.

Another approach, and in principle the optimal one, is to build a differentiable non-linear data model that starts from the Gaussian initial conditions and forward models them accurately to the measured lensing signal. This forward model is used to model the posterior of the parameters of interest (these can be cosmological parameters, the modes of the initial field or the bandpowers of the initial power spectrum). Analyzing this posterior generally relies on powerful sampling or optimization schemes, which in turn require many model evaluations and the derivatives of the model with respect to the parameters of interest. Forward-model based inference schemes have been developed for a range of of observables in cosmology~\citep{Seljak1998, Seljak17, Borg1, Borg2} including weak cosmic shear~\citep{Boehm17,Porqueres20}. 

All of these new avenues for lensing analyses create the need for fast and differentiable simulations of the lensing field that at the same time accurately capture the nonlinear features of that field.

Realistic lensing simulations are challenging because a range of scales in the three dimensional matter distribution contribute to a single angular scale in the lensing field making even intermediate lensing scales sensitive to the non-linearity of structure formation on small scales~\citep{Jain97}. Accurate lensing simulations rely on lightcones constructed from high resolution N-body simulations. These N-body simulations must accurately resolve small scales, but must at the same time be large enough to produce lensing maps with an extent of several degrees.

A number of recent works have studied the applicability of deep generative models, in particular generative adversarial networks (GANs) for producing accurate lensing convergence maps at low computational costs~\citep{Mustafa19,Perraudin20}. These models do not aim at simulating the underlying physics, but are trained to mimic the training data to a degree where their output becomes indistinguishable from the training data for a neural network. While these early studies look promising, future research will have to show that these models indeed learn the correct data distribution or that using their outputs for inference leads to unbiased parameter posteriors. 
Another, more safeguard approach is to use machine-learning inspired techniques to boost the accuracy of low resolution N-body simulations~\citep{Dai20,DaiSeljak20} and to construct lightcones from those. This is the avenue we have chosen in this work to create high resolution lensing simulations from approximate N-body solvers.

In this publication we describe a new, weak gravitational lensing package, MADLens. MADLens is a python package that allows to compute fully nonlinear lensing convergence maps at different source redshifts and low computational cost while accurately modeling the non-Gaussianity of the field down to scales of several tens of arcseconds. MADLens is built on top of a particle-mesh solver that evolves an initial linear density field into non-linear late time density fields. It provides derivatives with respect to the initial conditions of the particle-mesh simulation and a number of cosmological parameters through automated differentiation. MADLens fills the gap in the accuracy-speed space between computationally expensive, high accuracy lensing simulations and fast approximate simulations. In particular, MADLens correctly captures scales down to $L{=}10000$ at a field of view (FOV) of $6.2^\circ$ with percent level precision at a runtime of 30 seconds on 32 processes. MADLens can be run at different levels of resolution and approximations, allowing the user to choose the speed to accuracy trade-off that is optimal for their application. 

We begin this paper with a brief introduction of the cosmological-scale weak lensing formalism and our notation in Section~\ref{sec:notations}. This is followed by a in depth discussion of the package design in Section~\ref{sec:design}. We demonstrate the packages abilities in a number of tests in Section~\ref{sec:res} and conclude with a summary and outlook in Section~\ref{sec:outlook}. \ref{sec:app1} provides details on the novel feature of differentiability with respect to cosmological parameters.

\section{Weak Gravitational Lensing, Notation and Conventions}
\label{sec:notations}
Weak gravitational lensing observations provide insight into the projected matter density distribution between an observer and a source through correlated image distortions.  Here, we provide a brief overview and define our usage of the lensing kernel, lensing convergence, and power spectrum of cosmic shear used throughout MADLens (for a detailed discussion of weak gravitational lensing and especially cosmic shear, see~\citet{Bartelmann2017, Kilbinger2015, Bartelmann1999}). 

The image of a source galaxy at a comoving distance $\chi_s$ is distorted along the line of sight by some lensing potential $\Psi$. The potential of an extended lens under the Born approximation representing all density fluctuations along a line of sight at some angular position $\Vec{\theta}$ can be found by integrating individual Weyl potentials $\Phi$ up to the comoving distance of the galaxy, 
\begin{equation}
    \begin{split}
        \Psi(\Vec{\theta}) = \dfrac{2}{c^2}\int_0^{\chi_s} d\chi \dfrac{\chi_s-\chi}{\chi_s\chi}\Phi(\chi_{\Vec{\theta}}, \chi),
    \end{split}
\end{equation}
where $\chi_{\Vec{\theta}}$ is the angular perpendicular component of the potential and $\chi$ is the parallel component. The lensing deflection and convergence are defined as, 
\begin{equation}
    \begin{split}
        \Vec{\alpha} = \Vec{\nabla}\Psi, \\
        2\kappa = \nabla^2\Psi,
    \end{split}
\end{equation}
where the derivatives are taken with respect to $\vec{\theta}$.  Using the Poisson equation and neglecting derivative along the line-of-sight direction, the lensing convergence can be rewritten as, 
\begin{align}
\label{eq:conv}
    \kappa(\theta) &= \dfrac{1}{2}\nabla^2\Psi(\theta) = \dfrac{3H_0^2\Omega_{m0}}{2c^2}\int_0^{\chi_{\mathrm{lim}}} \dfrac{d\chi}{a(\chi)}q(\chi)\delta(\chi\theta, \chi)\\
    q(\chi) &= \int_\chi^{\chi_{\mathrm{lim}}}d\chi' n(\chi')\dfrac{\chi(\chi'-\chi)}{\chi'},
\end{align}
where $\delta$ represents the density contrast from the mean density, $H_0$ is the Hubble parameter and $q$ is the lensing kernel describing the projection of sources selected by the redshift selection function $n(\chi')$. MADLens evaluates the integral in Eq.~\ref{eq:conv} numerically.
The most commonly used summary statistic in lensing analyses that can also be computed analytically is the power spectrum. To compute the convergence power spectrum we use Limber's approximation and the flat sky approximation which are both valid on intermediate and small scales,
\begin{equation}
\label{eq:los}
C^{\kappa\kappa}_L = \left[\frac{3H_0^2\Omega_{m0}}{2c^2}\right]^2 \int_0^{\chi_s} d\chi \left(\frac{\chi_s-\chi}{\chi_s}\right)^2 P_m\left(k=\frac{l+0.5}{\chi}, z(\chi)\right).
\end{equation}
Since the lensing power spectrum is very sensitive to non-linear corrections to the matter power spectrum, we use HaloFit~\citep{Takahashi2012} to model $P_m(k)$ throughout this paper.
\begin{table*}
    \centering
    \begin{tabular}{l|l|l}
        Parameter &  Description & Typical Value(s) \\
        \hline
        BoxSize & side length of the simulation box & 128-1024 Mpc/h \\
        Nmesh & resolution of the particle-mesh simulation & $64^3-512^3$ \\
        B     & force resolution factor & $2$ \\
        Nsteps & number of steps in the FastPM simulation & $11-40$ \\
        N\_maps  & number of output maps & $\geq 1$ \\
        Nmesh2D & resolution of the convergence map & $256^2-2048^2$\\
        BoxSize2D & size of the convergence map in degrees & $2.5^\circ-22^\circ$\\
        zs\_source & list of source redshifts & $0.3-2.0$ \\
        Omega\_m & total matter density & $0.32$ \\
        sigma\_8 & amplitude of matter fluctuations & $0.82$ \\
        PGD & whether to use PGD enhancement or not & True/False \\
        interpolation & whether to use the sub-evolution scheme & True/False 
    \end{tabular}
    \caption{\label{tab:MADLens_params} List of MADLens simulation parameters that can be set by the user and their typical values.}
    \label{tab:my_label}
\end{table*}

\section{MADLens package design}
\label{sec:design}
The MADLens package is based on FastPM~\citep{fastpm}, a highly scalable particle-mesh solver, that evolves particle positions through a kick and drift scheme enforcing correct linear displacement in each step. FastPM has been implemented in C and Python and two versions of FastPM support automatic differentiation, including the MPI based version used in this work, and the recently published FlowPM~\citep{Modi20} package, which is based on TensorFlow. A FastPM particle-mesh simulation requires the choice of a particle-mesh resolution, equivalent to the number of particles in the simulation, the force resolution, the resolution of the grid onto which the particles are painted to compute the forces, the box size of the simulation and the number of steps in which the particle positions are evolved. The initial conditions and particle evolution depend on the cosmological parameters $\sigma_8$ and $\Omega_{m0}$.

MADLens runs a FastPM simulation. As the simulation evolves, MADLens projects the particles in the simulation weighted by the lensing kernel at each simulation step to 2D meshes at the desired source redshifts. The field of view, i.e. the size of the convergence map, and its resolution can be set by the user given the simulation box covers the entire field of view at the most distant source redshift. 
Each FastPM step evolves particles in redshift steps, $\Delta z_i=z_{i+1}{-}z_i$. MADLens constructs the lightcone by translating these redshifts into distances $\Delta \chi_i = \chi_{i+1}{-}\chi_i$ and projecting particles at the correct evolution step corresponding to that distance onto the convergence map. 
If the distance between two simulation steps is larger than the extent of the box, the box is replicated at the same redshift as often as is needed to fill the entire extent. In order to avoid spurious correlations, the simulations box is rotated before being repeated.
In these techniques, MADLens is constructed similarly to other lightcone packages. We provide an overview of all MADLens parameters that can be set by the user in table~\ref{tab:MADLens_params}. 

To reach extraordinary accuracy at low computational costs, MADLens employs two special techniques: 
\begin{itemize}
\item Particle Gradient Descent (PGD)~\citep{PGD} is an additional particle evolution step that corrects for the difference between particle distributions in a low resolution simulation and a high resolution simulation. The correction is applied after each simulation step. PGD introduces 5 additional nuisance parameters, which are fitted on training simulations. PGD allows simulations to run at lower resolution while still obtaining results that are comparable and highly correlated with a high resolution simulation.
\item A sub-evolution step allows for a massive reduction of the number of simulations steps. When using sub-evolution, particles are evolved according to the redshift of their position within the simulation box before projection, rather than by the redshift of the FastPM step.
\end{itemize}
Finally, MADLens is made differentiable through numerically accurate tape-based automatic differentiation. Specifically it uses the Virtual Machine Automated Differentiation package (VMAD). VMAD builds a graph that is traversed for the model evaluation, during which all operations are recorded on a sequential tape. Gradient graphs generated from the tape are used to compute Jacobian vector products ($J_{ij} v_j$) and vector Jacobian products ($v_i J_{ij}$, commonly referred to as back-propagation). 
MADLens is made available in two variants. In its main version it is built to provide differentiability with respect to the initial, Gaussian modes of the simulation. A second package version, that is included in this release, adds differentiability with respect to the cosmological parameters, $\Omega_{m0}$ and $\sigma_8$. Differentiability with respect to the PGD paramaters ($k_l$, $k_s$, $\alpha_0$, $\mu$) will be included in a future release.
\section{Results}
\label{sec:res}
We analyze the performance of MADLens, the PGD enhancement, the sub-evolution scheme and computation times as well as the accuracy of the gradient computation. 
Testing the accuracy of MADLens output is challenging, because of the lack of a ground truth. We will use theoretical power spectra based on Halofit matter power spectra and high resolution runs for comparison. These can serve as a reasonable baseline, but as should become evident from our analysis, should not be mistaken for the ground truth.

\begin{figure}
\centering 
\includegraphics[width=0.8\columnwidth]{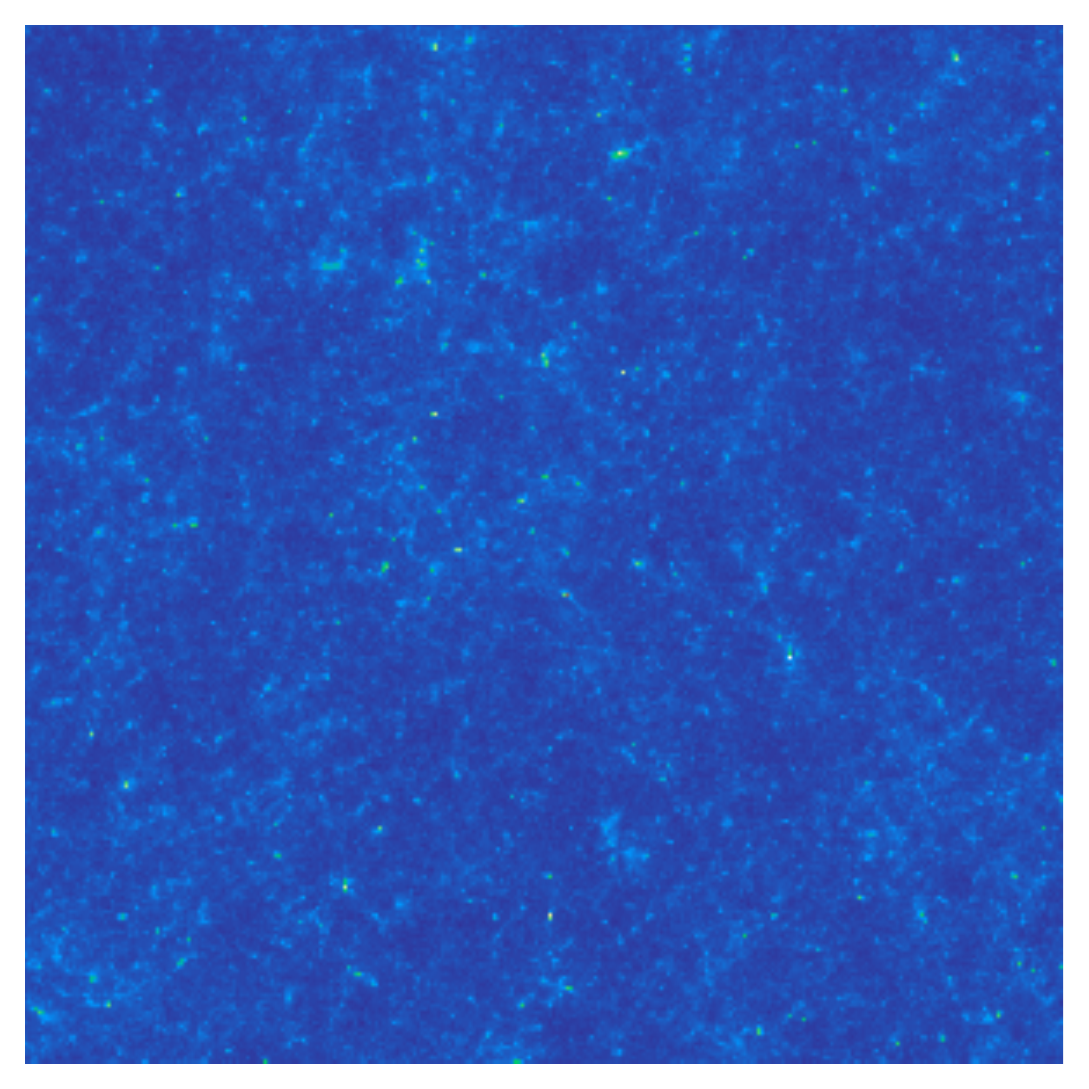}
\caption{\label{fig:map} A MADLens convergence map at $z{=}1.0$, based on a 3D simulation of side length $256~\mathrm{Mpc/h}$ and $256^3$ particles. The 2D lensing map has an angular extent of $6.2^\circ$. It was down-sampled to a map of pixel size $43~\mathrm{arcsec}$ and Nyquist frequency $L{=}15000$. The Non-Gaussianity is clearly visible by eye.}
\end{figure}
\begin{figure}
\centering 
\includegraphics[width=0.99\columnwidth]{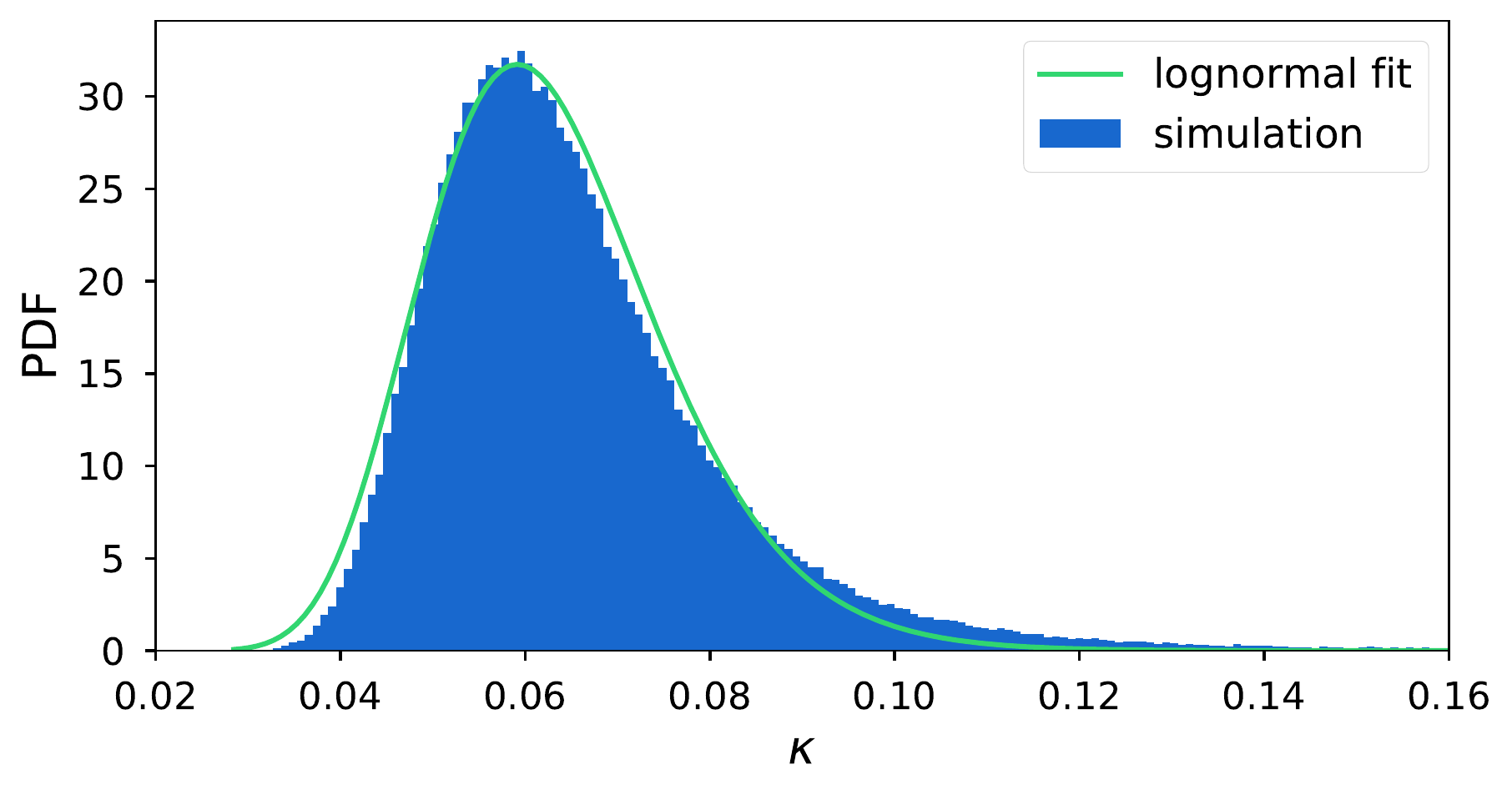}
\caption{\label{fig:hist} PDF of convergence values in the map in Figure~\ref{fig:map}. The distribution is clearly non-Gaussian with a pronounced skewness. A lognormal $\chi^2$ fit to the distribution (green) approximates its shape to some extent but does not accurately capture the tails.}
\end{figure}

In Figure~\ref{fig:map} we show an example of a convergence map produced with MADLens. The non-Gaussian structure is clearly visible by eye and also evident in the corresponding histogram in Figure~\ref{fig:hist}. The map resolves sub arcmin scales and extends over more than $6^\circ$ on the sky showing that MADLens overcomes one of the key challenges in lensing simulations: the accurate modeling of both large and small scales.

We compare the histogram of pixel values in Figure~\ref{fig:hist} with a log-normal distribution. Log-normal distributions have been used in the past to model lensing PDFs, but are not strictly theoretically motivated. The log-normal fit captures the rough shape of the distribution, but underestimates the probability of high values and overestimates the probability of low values. 

\subsection{Accuracy}
\label{sec:res_a}
\begin{figure}
\centering 
\includegraphics[width=0.99\columnwidth]{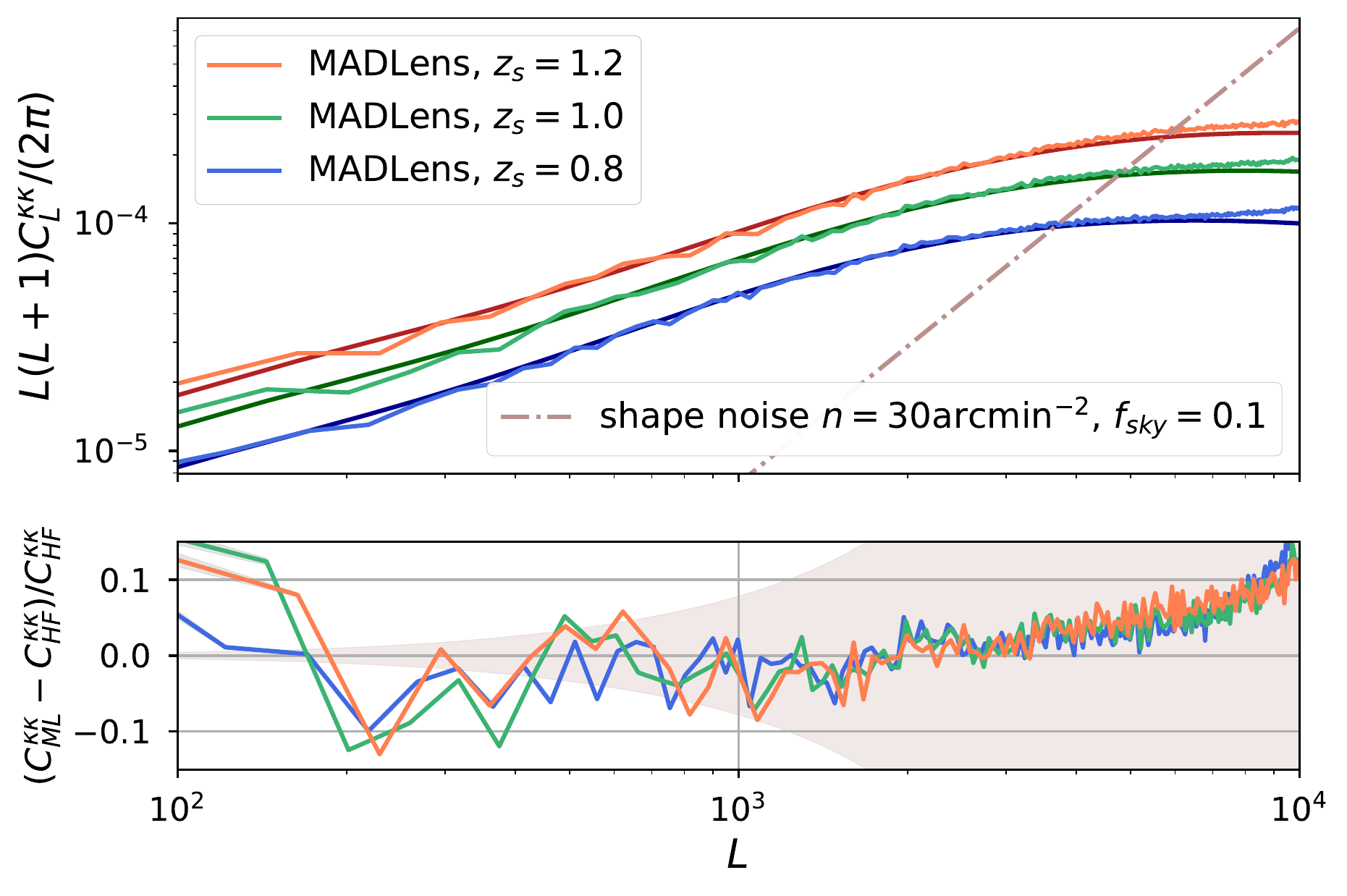}
\caption{\label{fig:zsource}  MADLens outputs (box length 512 Mpc/h, $512^3$ particles, with sub-evolution and PGD enhancement) for different source redshifts but same initial conditions (no shot noise subtraction). The measured power spectra agree with theoretical predictions up to very high wavenumbers independent of the source redshift. Lower source redshifts show slightly higher shot noise due to the lower number of particles that contribute to the projection. For comparison we plot the experimental noise for a typical galaxy density expected for future lensing experiments (pink dashed-dotted line and shaded area), showing that areas with significant shot noise contribution lie well within the experimental noise dominated regime.}
\end{figure}
In Figure~\ref{fig:zsource} we compare the power spectra measured from MADLens outputs at different source redshifts with the analytical model of Eq.~\ref{eq:los} based on a HaloFit matter power spectrum. For this comparison we average the power spectra of five simulations to reduce the variance. Overall we find that the MADLens power spectra trace the theoretical predictions well within $10\%$ up to scales of a few thousand. At very small scales shot noise starts to contribute significantly to the power. To put the importance of this shot noise into perspective we further plot the experimental noise level expected in a typical future experiment, such as LSST, and find that the shot noise is subdominant to the expected noise levels in real data. 

\begin{figure}
\centering 
\includegraphics[width=0.99\columnwidth]{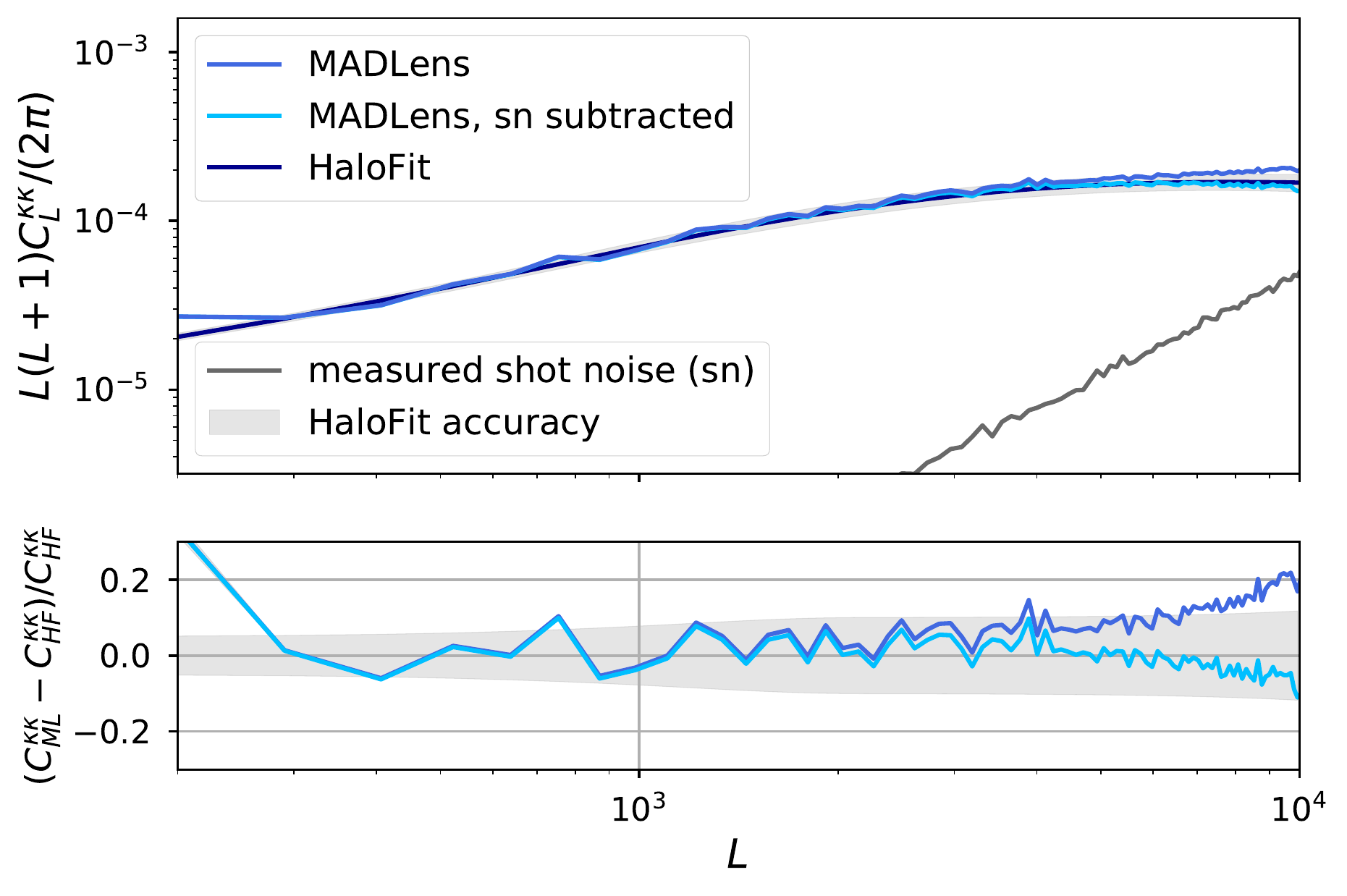}
\caption{\label{fig:accuracy} Shot noise and comparison to a theoretical convergence power spectrum based on a HaloFit matter power spectrum: The MADLens simulation traces the HaloFit power spectrum within the accuracy of HaloFit (gray band). At high wavenumbers ($L{>}4000$) the shot noise (dark gray line) that is due to the small number of particles in the simulation ($256^3$) starts to contribute significantly to the signal. After subtraction of the shot noise, the convergence power lies well within the uncertainty band up to $L{=}10000$.}
\end{figure}
Figure~\ref{fig:accuracy} delves further into the comparison with theoretical power spectrum and quantification of the shot noise. We translate the HaloFit accuracy ($5\%$ for $k\leq1~\mathrm{h\,Mpc}^{-1}$ at $0\leq z \leq 10$ and $10\%$ for $1\leq k\leq10~\mathrm{h\,Mpc}^{-1}$ at $0\leq z\leq 3$) into accuracy in the lensing power spectrum and show these intervals as gray bands. The MADLens power spectrum lies well within these bands up to wavenumbers of a few thousand, where it becomes dominated by shot noise. We estimate the shot noise level by running a number of MADlens simulations with random particle positions (dark gray line) and subtract the result from the MADLens power spectrum (dark blue). The result lies within the HaloFit accuracy up to $L{=}10000$.

\begin{figure}
\centering 
\includegraphics[width=0.99\columnwidth]{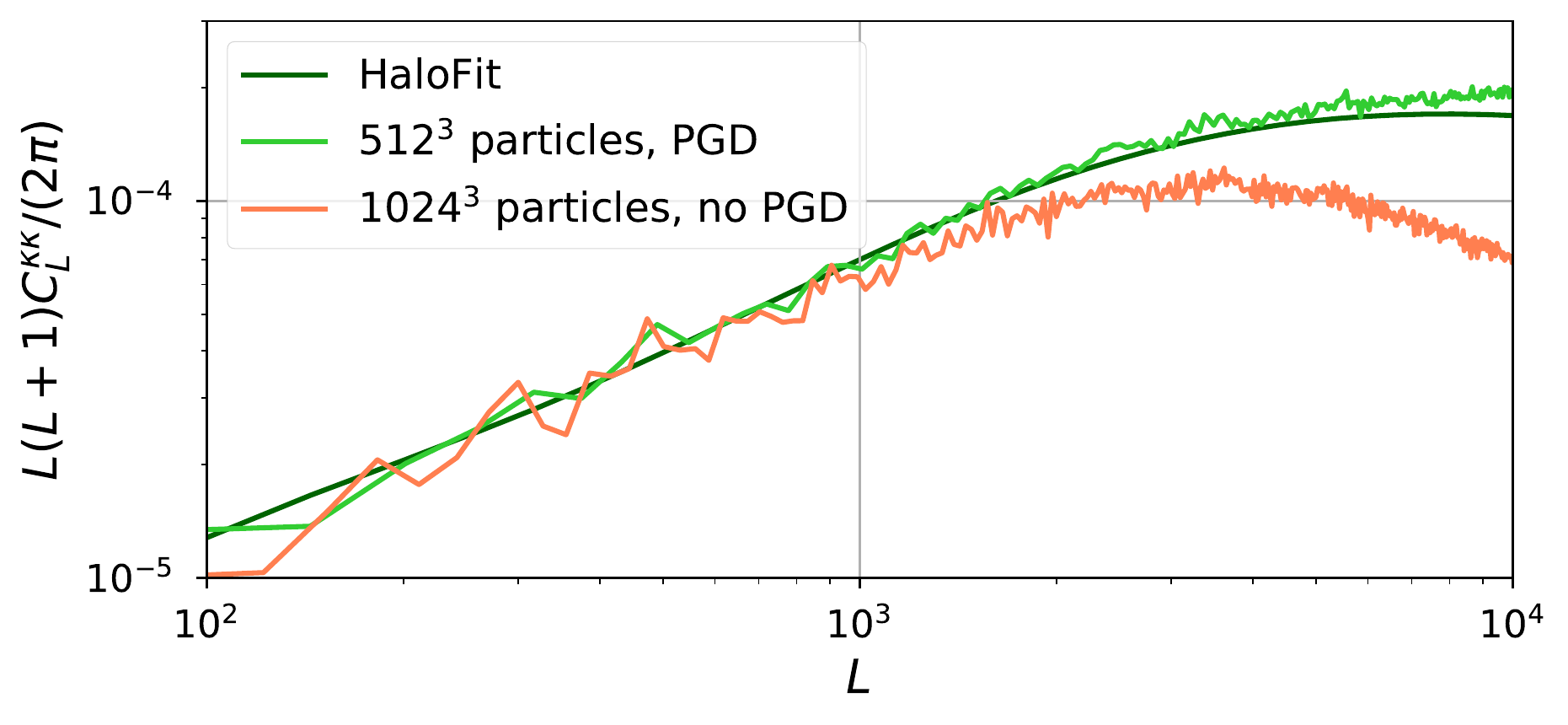}
\caption{\label{fig:PGD} The PGD enhancement recovers lensing power on small scales in a low resolution simulation: We show a comparison between the convergence power spectra of a MADLens simulation with $512^3$ particles in a box of side length 512 Mpc/h that was run with PGD enhancement (light green) and an otherwise identical simulation with 8 times more particles that was run without PGD (orange). The latter corresponds to a setting used in some state-of-the-art simulations. The PGD enhanced simulation tracks the theory power spectrum (dark green) well beyond $L{=}1000$, while the standard simulation suffers from a significant loss in power on small scales.}
\end{figure}
Figure~\ref{fig:PGD} shows that the PGD enhancement allows to reach these high accuracies at much lower computational cost than conventional lensing simulations. We compare the output of MADLens simulations at a resolution of 1 particle per Mpc/h cubed to a conventional simulation (MADLens without PGD) with an eight times higher resolution. The higher resolution simulation not only requires about eight times more memory, but also takes more than twice as long. The high resolution run is of comparable resolution to other state-of-the-art lensing simulations which have been used for cosmological parameter inference studies~\citep{Liu18}, but the lower resolution MADLens simulation traces the theoretical convergence power spectrum up to much higher wavenumbers.

\begin{figure}
\centering 
\includegraphics[width=0.99\columnwidth]{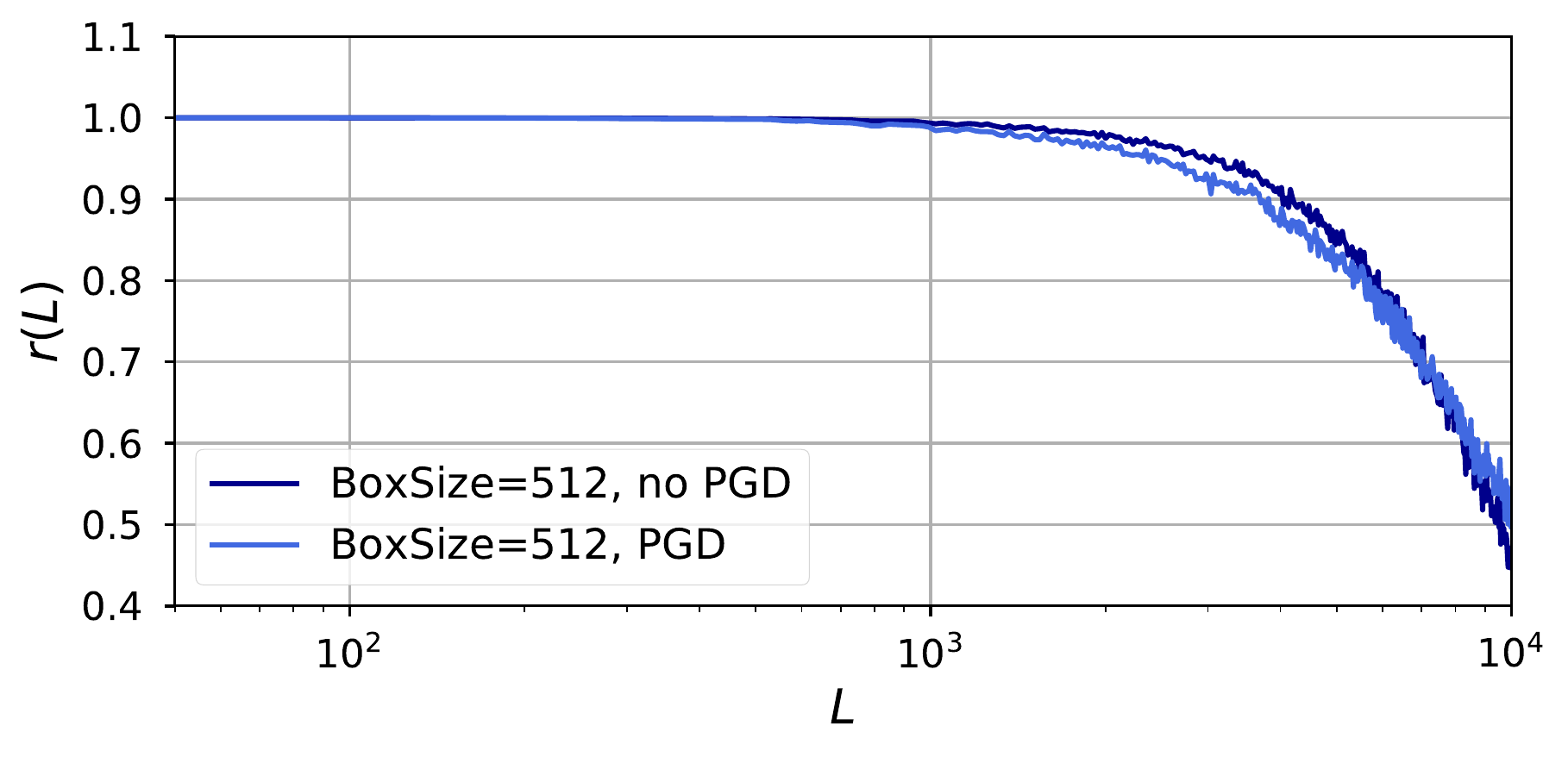}
\caption{\label{fig:cross_corr} Cross correlation between a low resolution MADLens run with and without PGD enhancement, and a simulation with 8 times more particles and no PGD enhancement (same as Fig~\ref{fig:PGD}). The PGD enhancement increases the cross correlation on intermediate scales demonstrating that PGD moves particles in a physically sensible way. Note that the high resolution simulation is not the \textit{truth} as it lacks a significant amount of power on small scales.}
\end{figure}
In Figure~\ref{fig:cross_corr} we show cross correlations defined by
\begin{equation}
    r(L) = \frac{C_L^{XY}}{\sqrt{C_L^{XX}C_L^{YY}}},
\end{equation}
where $X$ is the high resolution run without PGD enhancement, and $Y$ are lower resolution MADLens outputs that have either been produced with or without PGD enhancement. As expected, the PGD enhanced lower resolution map shows higher correlation (dark blue) with the high resolution run than the one without enhancement (light blue) on intermediate scales. The results of this cross correlation analysis must be taken with a grain of salt: since the high resolution run is suffering from a significant lack of power on small scales, this could also be an indication of inaccurate particle positions. A lower cross correlation on small scales simply states that the simulations differ significantly, but does not show which one is more correct.

\begin{figure}
\centering 
\includegraphics[width=0.99\columnwidth]{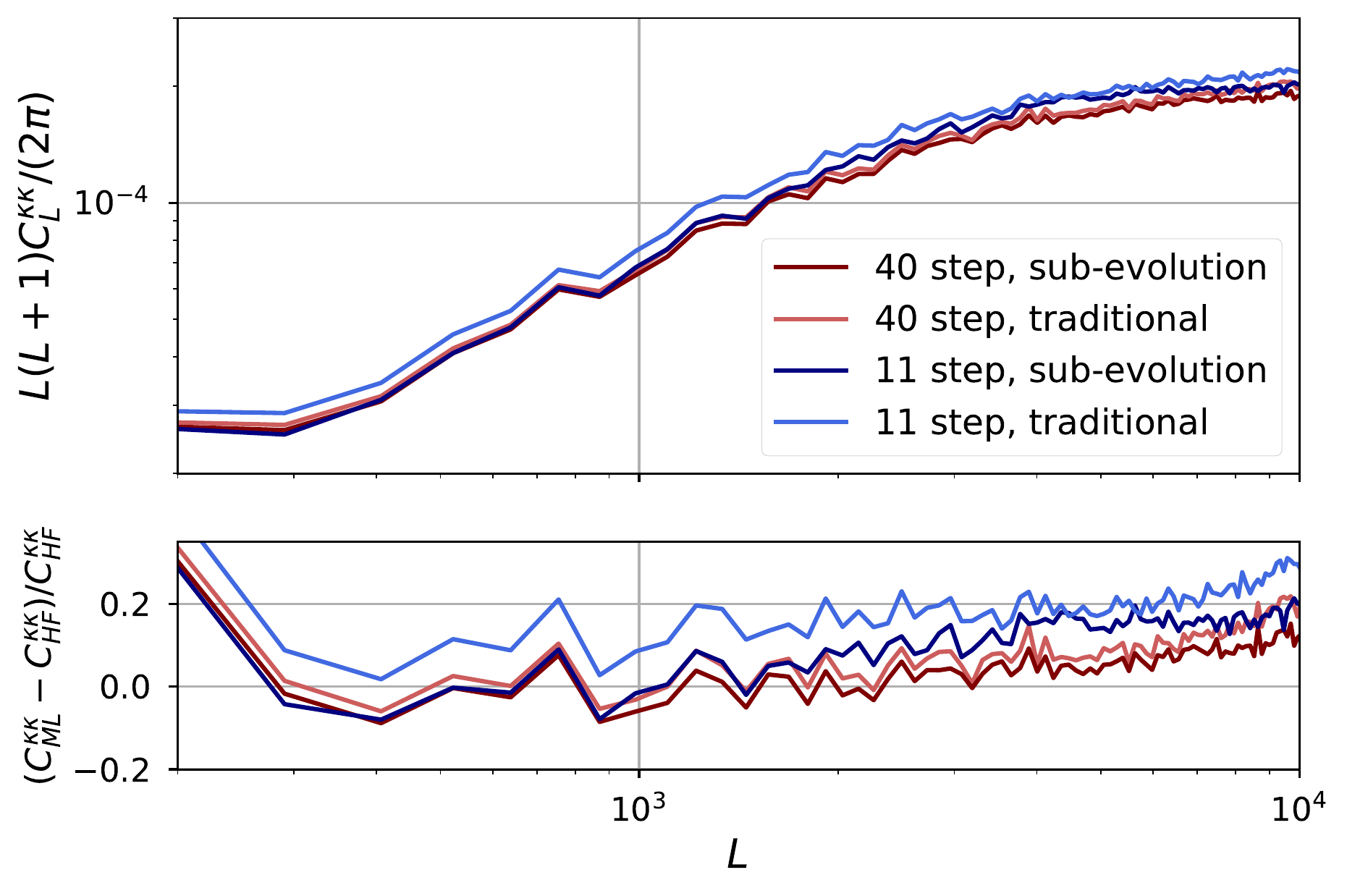}
\caption{\label{fig:sub_evolution} The sub-evolution scheme corrects for a systematic overestimation in power that occurs when snapshots are projected at the end of every FastPM step instead of at the exact position/redshift that corresponds to their distance to the observer (light blue). While the discrepancy can be reduced by using more simulation steps and hence higher computational cost (light red), a sub-evolution step before each projection significantly improves the output at a fixed number of simulation steps (dark blue and red).}
\end{figure}
The accuracy of MADLens is further boosted by a sub-evolution projection scheme, where particles are moved to the position corresponding to their actual distance to the observer before being projected on the lensing map. We illustrate the efficacy of this scheme in Figure~\ref{fig:sub_evolution}. An 11-step simulation naturally overestimates the total lensing power (light blue). The sub-evolution scheme is able to correct for this overestimation up to scales $L{>}2000$ (dark blue). In a 40 step simulation the discrepancy between actual particle positions and their true evolution stage is smaller (light red), however, even the accuracy of a 40 step simulation can be enhanced by the sub-evolution scheme (dark red).

\begin{figure*}
\centering 
\includegraphics[width=0.95\textwidth]{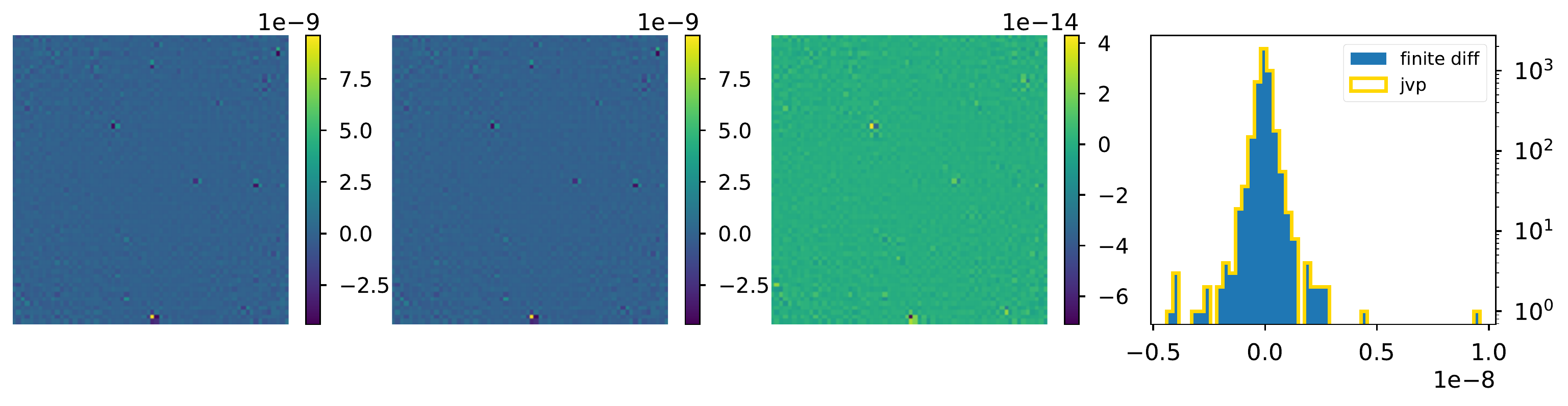}
\caption{\label{fig:vjp} The MADLens automated derivative agrees excellently with the result of finite differencing. To show this here, we measure the response of the convergence map by slightly changing the initial field in a single pixel. The corresponding MADLens Jacobian-vector-product in the first plot and the finite difference result in the second plot agree to the order of $10^{-5}$, as can be seen in the difference map (third plot) and by comparing their histograms (fourth plot).}
\end{figure*}
The MADLens derivatives have been thoroughly tested and verified with VMAD built-in test functions. Here we show that the derivatives are accurate by means of a single example: we build a finite difference test by slightly changing a single pixel value in the initial field. We then take the difference of the output maps generated from runs with slightly different values in this initial pixel and compare it with the output of the Jacobian-vector-product (Jvp), where the vector encapsulates the change in the initial field. If the Jvp vector product is correct, the result should agree with the difference of the output maps. That this is indeed the case is shown in Figure~\ref{fig:vjp}. The first two panels show the Jvp and the finite difference result, respectively. They are indistinguishable by eye. The next panel shows the difference between the first two panels, revealing insignificant numerical inaccuracies, five magnitudes smaller than the signal. In the last panel we compare the outputs in terms of their histograms, finding again an excellent match.  

\subsection{Run times}
\begin{figure}
\centering 
\includegraphics[width=0.99\columnwidth]{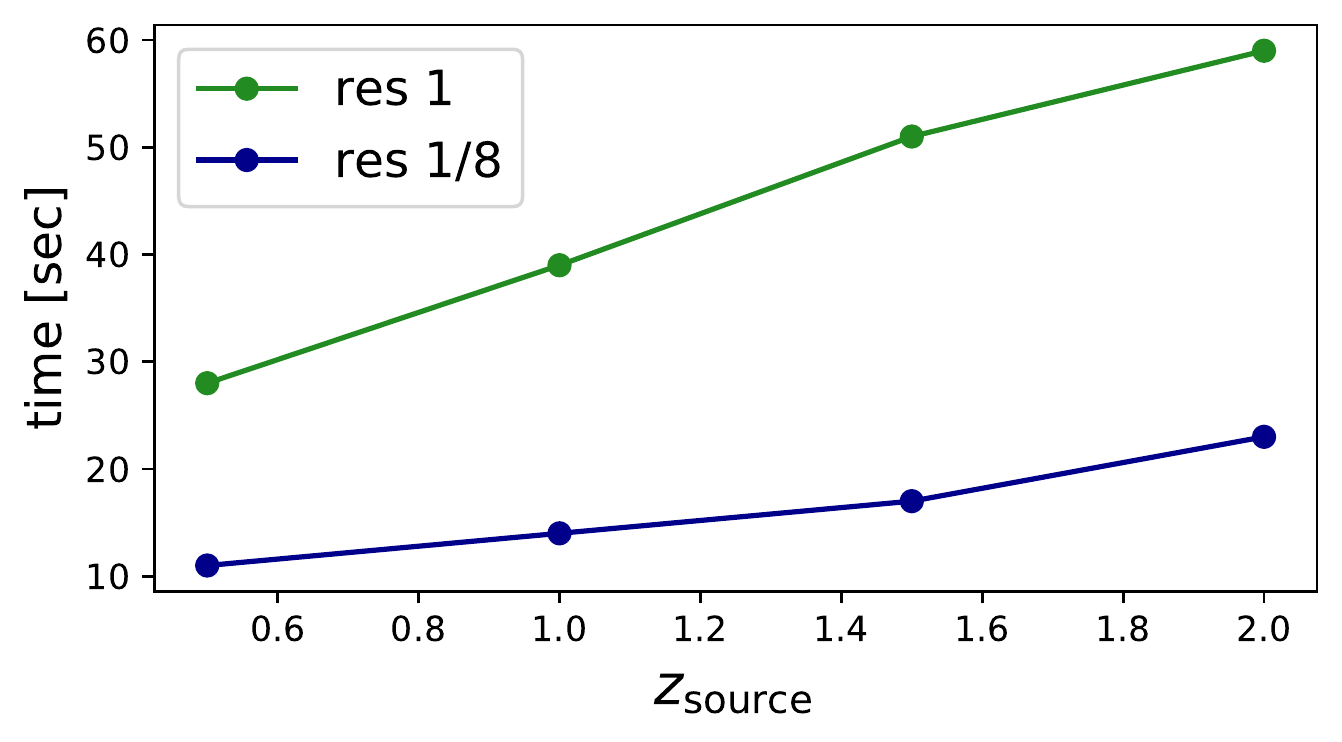}
\caption{\label{fig:timing} Computation time for a single simulation run with 11 steps and a box size of 256 Mpc/h for different source redshifts. We compare a run with $256^3$ particles ($\mathrm{res}=1\,\mathrm{[h/Mpc]}^3$) particles to a run with $128^3$ particles ($\mathrm{res}=0.125\,\mathrm{[h/Mpc]}^3$). All simulations were run on a single node with 32 processes in this test. }
\end{figure}

\begin{figure}
\centering 
\includegraphics[width=0.99\columnwidth]{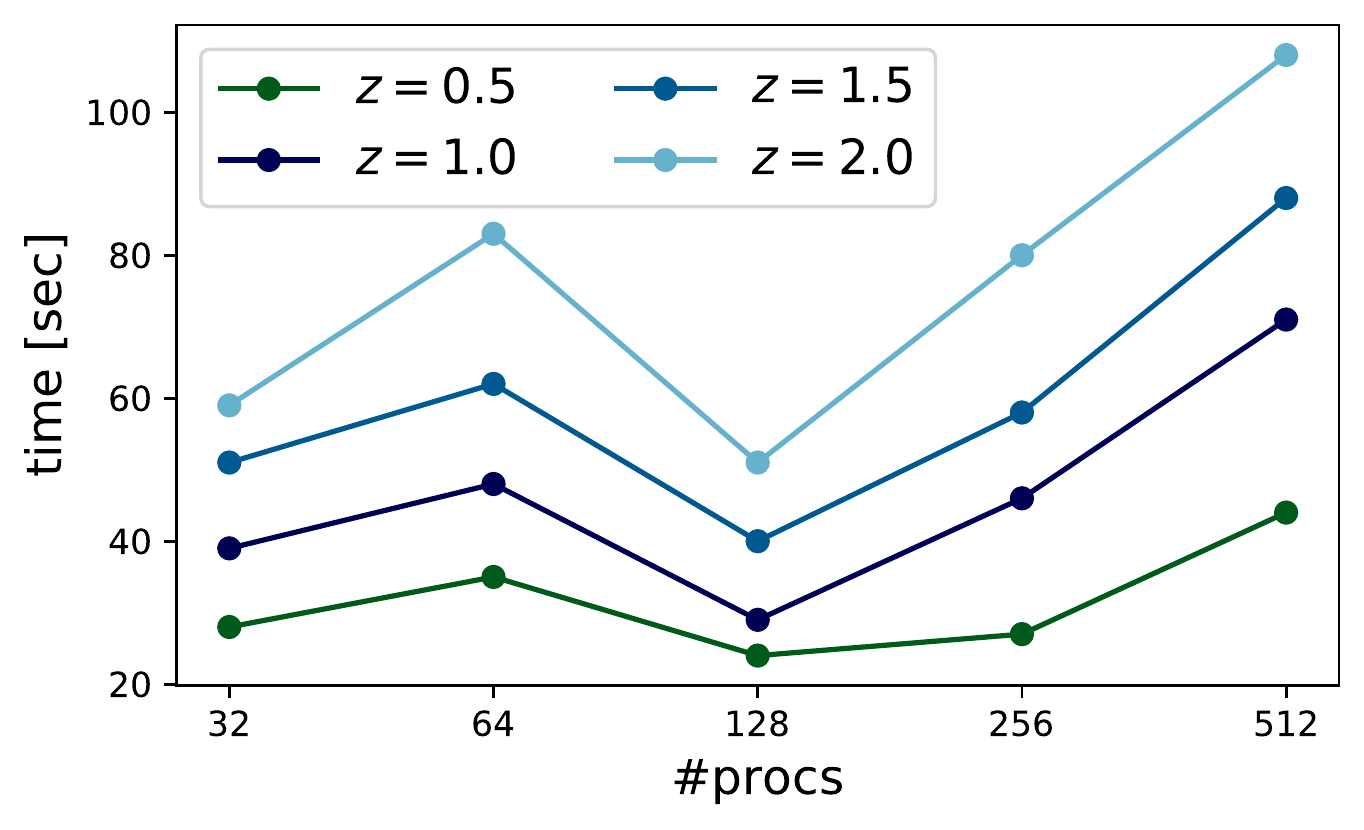}
\caption{\label{fig:timing_scaling} Scaling of the computation time with the number of processes for a $256^3$ particle/ 256 Mpc/h box simulation. Different lines represent different source redshifts.}
\end{figure}

\begin{figure}
\centering 
\includegraphics[width=0.99\columnwidth]{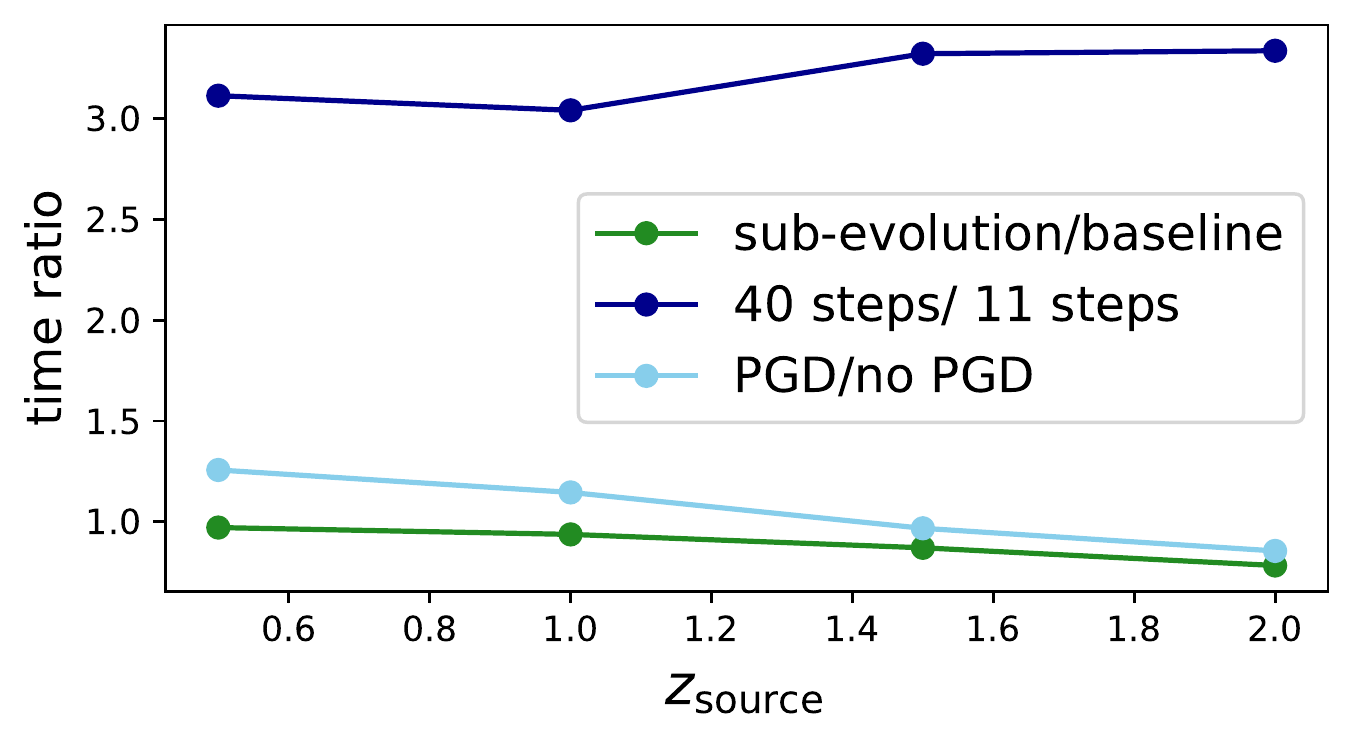}
\caption{\label{fig:ratio_timing} A comparison of different settings of the simulation package in terms of computation time. Using an 11 step simulation instead of a 40 step simulation reduces the computation time by more than a factor of 3. This reduction in computation time without paying a significant penalty in simulation accuracy is made possible by the PGD enhancement and the sub-evolution step. Neither of those enhancements change the computations times significantly (green and light blue lines). All simulations in this test were run with $256^3$ particles and a $256$ Mpc/h box.}
\end{figure}

\begin{figure}
\centering 
\includegraphics[width=0.99\columnwidth]{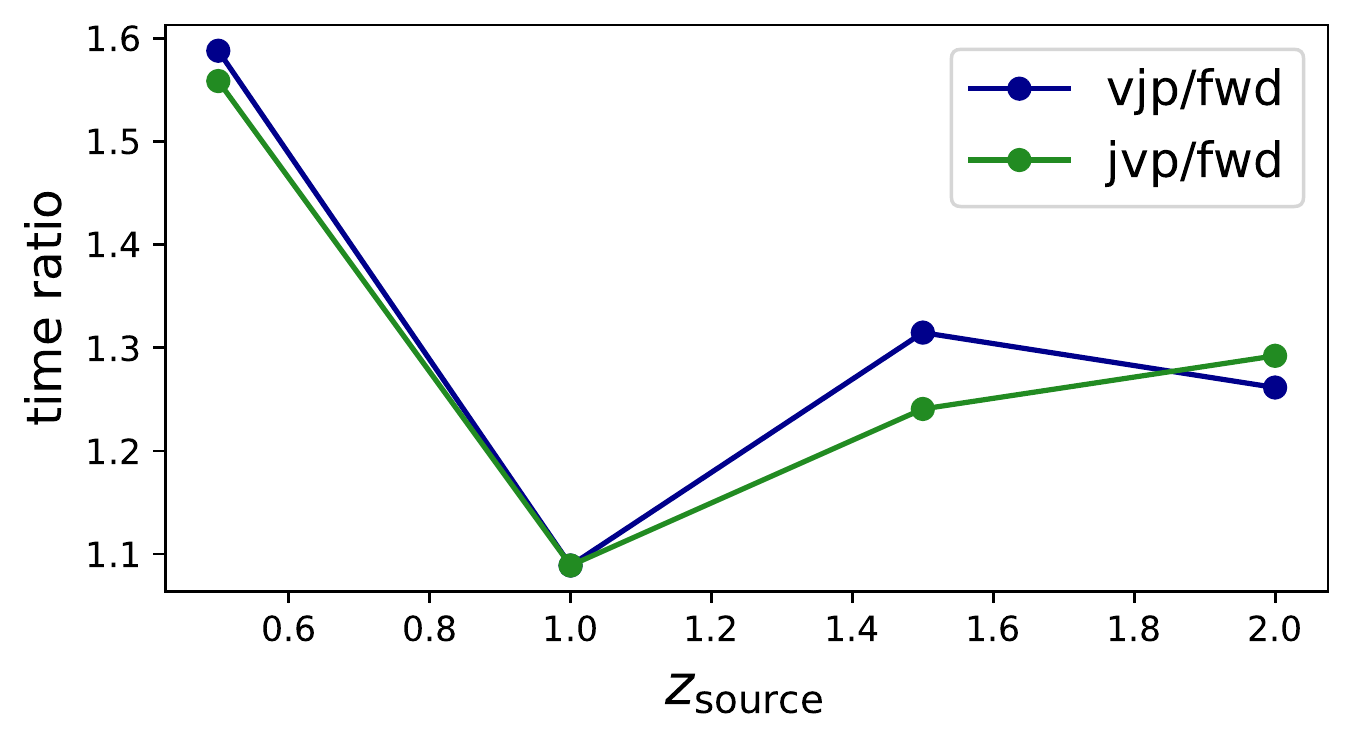}
\caption{\label{fig:deriv_timing} Computation time of vector-Jacobian (vJp) and and Jacobian-vector product (Jvp) in units of computation time of the forward model for different source redshifts. The vJp and Jvp require only slightly longer computation times than the forward model.}
\end{figure}

We conducted timing tests for MADLens on Intel® Xeon® Processors E5-2698 v3 (NERSC Cori Haswell nodes), and show results in Figures~\ref{fig:timing}-\ref{fig:deriv_timing}. A single MADLens simulation that achieves accuracies as shown in the last section takes of the order of 10-60 seconds on 32 processes. The scaling of the run-time with source redshift is roughly linear and reducing the number of particles by a factor of 8 reduces the run-time to about one third (Figure~\ref{fig:timing}).

The computation time can be further reduced by paralellizing on up to $128$ processes, after which the communication overhead starts to dominate the time budget (Figure~\ref{fig:timing_scaling}).

Reducing the number of FastPM steps leads to significant savings in run time as we demonstrate in Figure~\ref{fig:ratio_timing}. A conventional lightcone code requires about 40 steps in order to reach percent accuracies in the lensing power spectrum up $L{\approx}1000$. With PGD enhancement and sub-evolution scheme, MADLens reaches percent accuracies up to $L{=}2000$ with only 11 FastPM steps: a factor of 3 in time-savings. 

The use of back-propagation to calculate the derivatives results in run times that are similar to the forward model. This is shown in Figure~\ref{fig:deriv_timing}, where we find run times of 1.1-1.6 times the run time of the forward model for either Jvp and vJp.

\section{Summary \& Outlook}
\label{sec:outlook}
We have presented MADLens, a fully differentiable python package for producing non-Gaussian convergence maps of weak gravitational lensing on cosmological scales. MADLens reaches unprecedented accuracy even when compared to many non-differentiable lensing simulations, and operates at run times of the order of 2-20 below conventional N-body simulation based lightcone packages. These advancements are made possible by several features, including a machine learning inspired post processing step, that allows the N-body simulation to run at a lower resolution and with less steps without paying a significant penalty in accuracy. Taking the derivative through a MADLens simulation with respect to the initial modes of the N-body simulation and the two key cosmological parameters $\sigma_8$ and $\Omega_{m0}$ is made possible through back-propagation. This means that evaluating the derivatives has comparable computational cost as the forward simulation. 
With these features MADLens constitutes a milestone towards the development of fully differentiable inference pipelines for weak cosmic shear. 
In the future MADLens will be integrated into the tensorflow-based FlowPM framework. Package updates will also feature differentiability with respect to nuisance parameters, such as the PGD parameters. 

In the interest of scientific advancement and reproducibility, we make the MADLens package publicly available on github~\footnote{https://github.com/VMBoehm/MADLens}.  

\section*{Acknowledgements}
This research used resources of the National Energy Research Scientific Computing Center (NERSC), a U.S. Department of Energy Office of Science User Facility operated under Contract No. DE-AC02-05CH11231. The authors thank Uros Seljak for useful discussions and feedback and Fran{c}ois Lanusse for testing of the package and helpful feedback.

\appendix
\section{Differentiability with respect to Cosmological Parameters}
\label{sec:app1}
Since the forward model itself depends on cosmological parameters through the evolution of particle positions and the angular diameter distance which enters the lensing projection, an accurate inference algorithm needs to take these dependencies into account. To this end MADLens provides the additional functionality of derivatives with respect to the cosmological parameters $\Omega_{m0}$ and $\sigma_8$.  

This novel application of derivatives requires both power spectra, particle evolution, and comoving distance calculations written as functions of cosmological parameters.
The comoving distance calculation and derivative are trivial, and we use the standard definition \citep{Peebles93},

\begin{equation}
\begin{split}
    \chi = \dfrac{c}{H_0}\int_0^z \dfrac{dz'}{E(z')^{1/2}},
\end{split}
\end{equation}
where $E = \Omega_{m0}(1+z)^3 + \Omega_{k0}(1+z)^2 + \Omega_{\Lambda0}$. For a flat cosmology, we take $\Omega_{k0} \sim 0$ and $\Omega_{\lambda0} \sim 1-\Omega_{m0}$.

\begin{figure}
\centering 
\includegraphics[width=0.99\columnwidth]{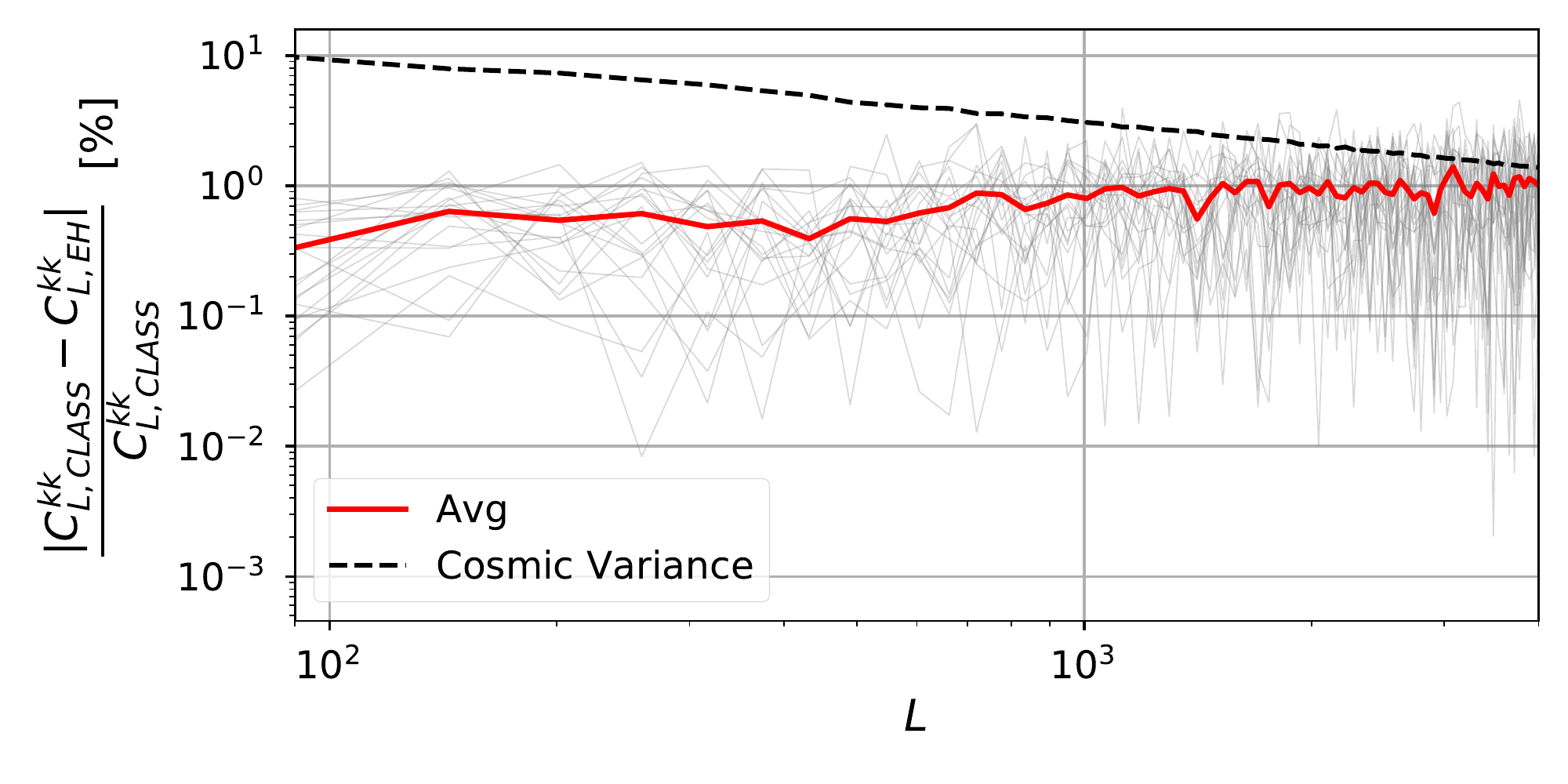}
\caption{\label{fig:EH_CLASS_Cl_comp} We calculate $20$ convergence maps using both CLASS and Eisenstein \& Hu transfer functions and see a strong agreement in values. The percent error of the power with respect to the CLASS transfer function convergence maps (red line) remains below the cosmic variance (dotted black line) for most modes with negligible deviations above at small scales.}
\end{figure}

For the power spectrum that is applied to the initial density field, we use the transfer function defined in~\citep{Hu99} (EH-Transfer) with the inclusion of baryonic acoustic oscillation (BAO) wiggles. This is much simpler and computationally less involved than obtaining gradients of standard Boltzmann solvers with respect to cosmological parameters. Compared to the the matter power spectrum obtained from the Boltzmann package, CLASS \citep{Blas11}, which is used for cosmological calculation throughout MADLens, we find discrepancies at a maximum of the $\sim5\%$ level. We find that by using the EH-transfer, we slightly overestimate power on all scales with he largest discrepancies at those corresponding to BAO wiggles. 

To show that this overestimation is within reason, we generate multiple convergence maps with both the CLASS and differentiable Eisenstein and Hu power spectrum. We show the absolute difference of the power spectra of these maps in Figure~\ref{fig:EH_CLASS_Cl_comp} and plot the cosmic variance for comparison. We find that the difference lies below the $1\sigma$ limit which we take to imply that our implementation of the EH-transfer function and power describes the initial spectrum well within the required accuracy. 

The particle initial conditions and evolution, too, is dependant on the cosmology and we use a finite differencing scheme on the Lagrangian Perturbation Theory initial conditions, as well as the momentum and position updates in FastPM. This allows the computation of approximate, and accurate, derivatives for the initial conditions, kick, and drift factors without the need for analytical solutions to the derivatives with respect to $\Omega_{m0}$. This method works by storing the finite difference of a function on the gradient tape, and caches the cosmology variables to avoid re-computation of the particle mesh. During forward propagation, the function is run as normal, and during back propagation two cached cosmology objects with perturbed parameter values are used to compute the finite difference.  This scheme can be made applicable to any function which is not highly sensitive to parameters, and while we use the analytic solution for derivatives such as the EH-transfer function, it should be noted that it is feasible to apply the finite differencing method to linear power and cosmology methods from Boltzmann codes such as CLASS. 

\begin{figure}
    \centering
    \includegraphics[width=.99\columnwidth]{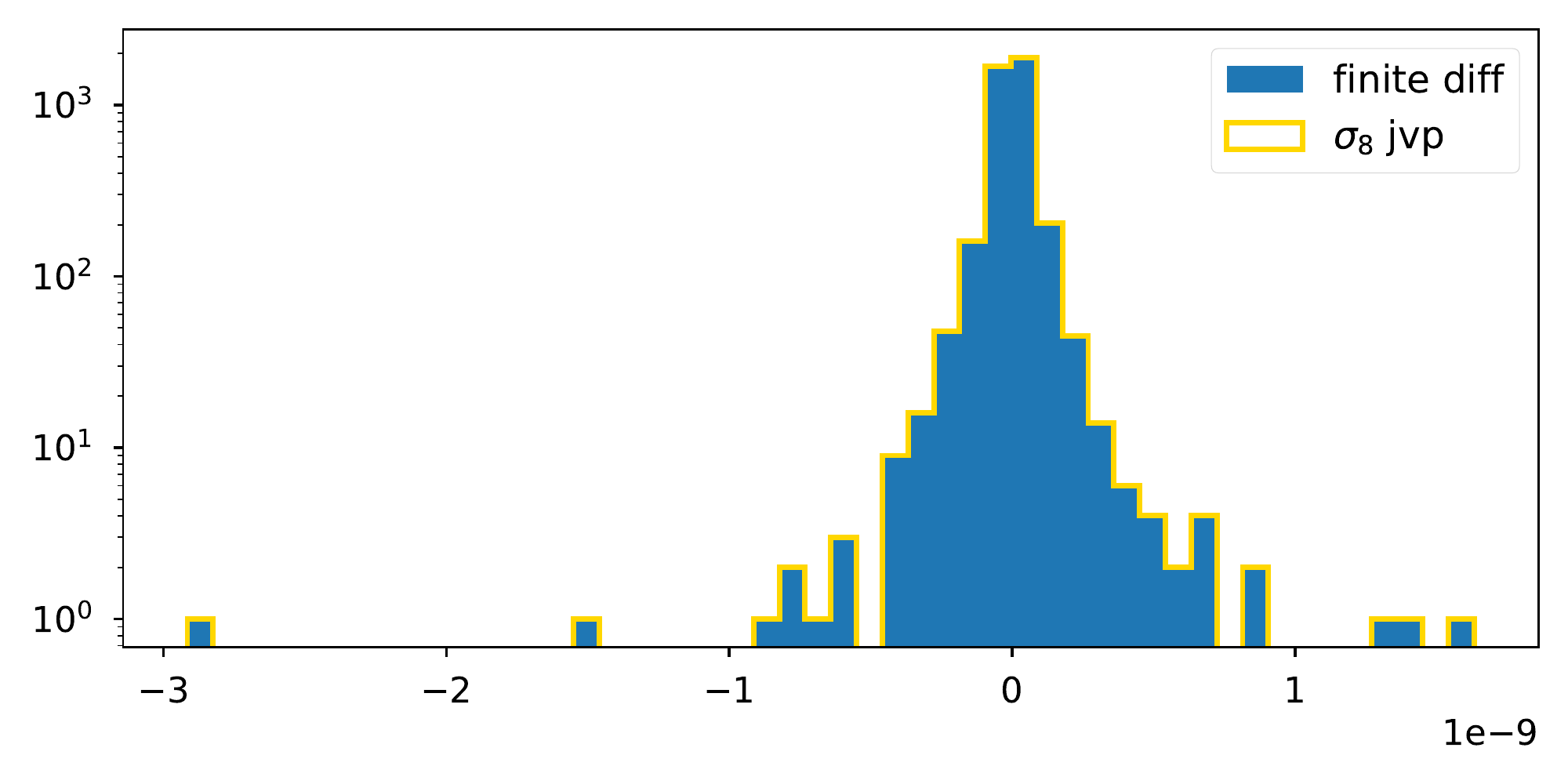}
    \caption{\label{fig:s8_derivs} We show that the automated derivative and finite differencing agree well for $\sigma_8$. We used a small offset $\delta =10^{-10}$ and histogram the finite differenced convergence maps (\textit{blue}) and Jvp outputs (\textit{gold})}
\end{figure}

Similar to the finite difference test in section~\ref{sec:res_a}, we test the accuracy of the Jvp for $\sigma_8$. We add and subtract a small offset $\delta$ from $\sigma_8$ ($\delta=10^{-10}$), compute convergence maps from both of these configurations and take their difference. We then compare this to the Jvp at the central $\sigma_8$ value with a vector of $2\delta$ to multiply the Jacobian. The resulting histograms are shown in Figure~\ref{fig:s8_derivs}. The derivative of $\sigma_8$ is in excellent agreement with the finite differencing result. 

\begin{figure}
    \centering
    \includegraphics[width=.99\columnwidth]{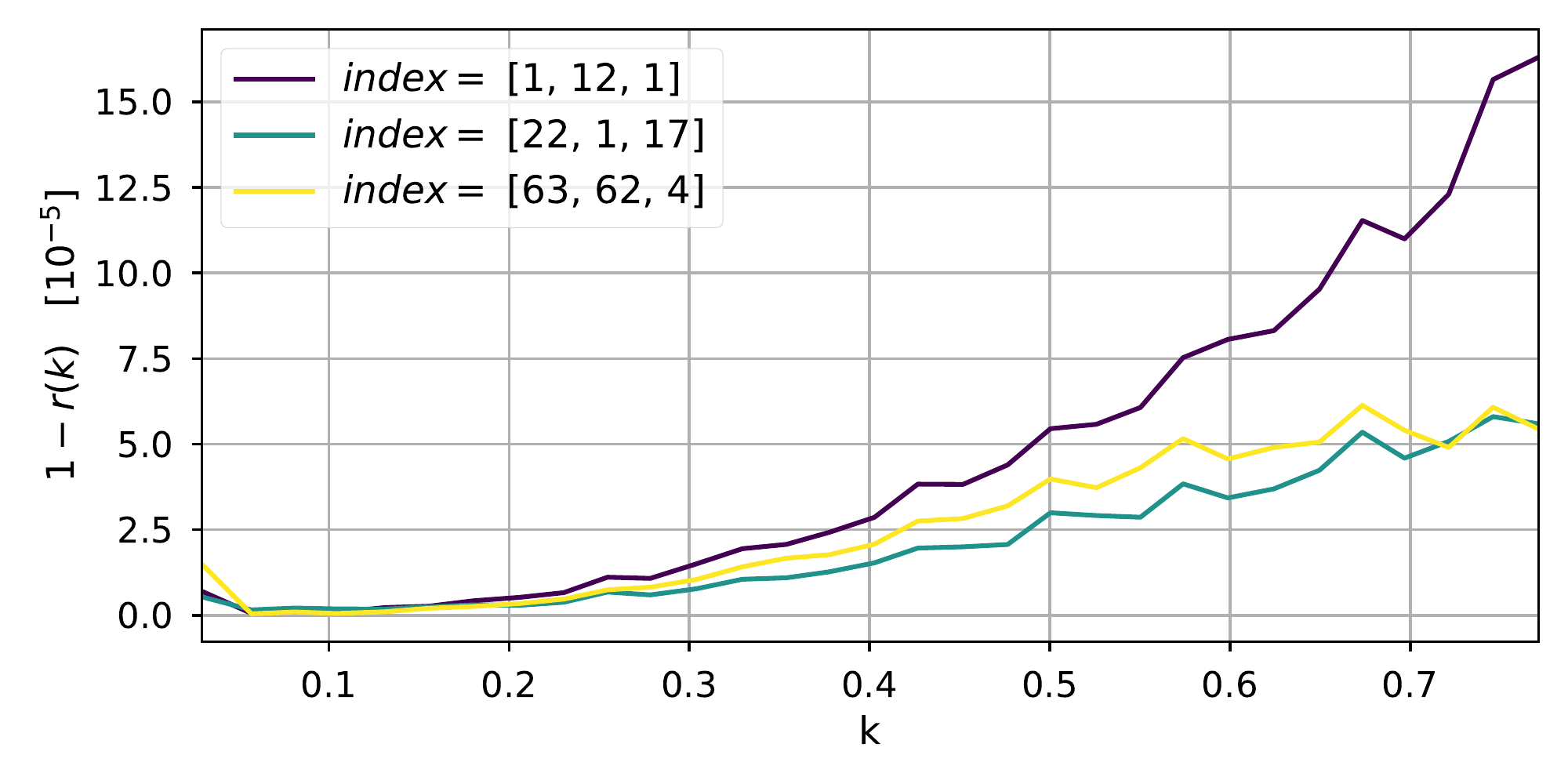}
\caption{\label{fig:Om0_corr} On each scale, the finite difference and automated derivative of $\Omega_{m0}$ agree to high accuracy. We show this by randomly selecting 3 modes to excite individually while setting all other modes to zero. In this plot, each line corresponds to the cross correlation between convergence maps generated when each of these selected modes is set to one, while the rest of the field is set to 0 for a $64^3$ mesh, and the corresponding Jvp. The indices used are specified in the legend. We find that irrespective of index choice, these correlations agree at the $10^{-5}$ level.}
\end{figure}

While $\sigma_8^2$ only enters linearly, the model's dependence on $\Omega_{m0}$ is more complicated. The derivative with respect to $\Omega_{m0}$ therefore requires a more in depth testing to ensure accurate derivatives are being taken. We test the Jacobian against finite differencing results for multiple modes individually. In Figure~\ref{fig:Om0_corr}, we show the results of cross-correlating the finite difference results with the Jvp outputs and verify the accuracy of the automated derivative. 

As a check for the vJp against $\Omega_{m0}$, we construct a scalar by computing the finite differencing of the sum of the squared convergence maps and ensure that this value is equal to the vJp when the central convergence map is used as the vector in automated differentiation. 
\begin{equation}
    \begin{split}
        \dfrac{\sum_{i=0}^{N} \kappa_{i, +\delta}^2 - \sum_{i=0}^{N} \kappa_{i, -\delta}^2}{2\delta} =2 vjp_{\Omega_{m0}}(v=\kappa) 
    \end{split}
\end{equation}
We find that the values agree at the $\sim 5\%$ level irrespective of the choice of $\delta$.


\bibliographystyle{plainnat}
\bibliography{ref}

\end{document}